\documentclass[12pt]{article}
\begin{document}

\renewcommand{\theequation}{\arabic{section}.\arabic{equation}}
 
\vspace{1.5 cm}

\noindent {\large\bf $CP^2$ SOLITON SCATTERING: 
THE COLLECTIVE COORDINATE APPROXIMATION} 
\vspace{1cm}

\noindent J. Burzlaff $\left.^{\dag}\:^{\ddag}\right.$
and W.J. Zakrzewski $\left.^* \right.$
\vspace{1cm}

\noindent $\dag$ School of Mathematical Sciences, Dublin City 
University, Dublin 9, Ireland.\\
\noindent $\ddag$ School of Theoretical Physics, Dublin Institute
for Advanced Studies, 10 Burlington Road, Dublin 4, Ireland.\\
\noindent $*$ Department of Mathematical Sciences, University of
Durham, Durham DH1 3LE, UK.
\vspace{1.5cm}

\noindent {\large\bf Abstract}
The $CP^2$ model, with and without a generalized Hopf term, is 
studied using the collective coordinate approximation. In the spirit of
this approximation, an ansatz is given which in previous numerical studies
was seen to give a good parameterization of the numerical solution. The 
equations of motion for the collective coordinates are then solved
analytically, for solitons close together and for solitons far apart. 
The solutions show how the generalized Hopf term changes the scattering 
angle which in its absence is $90^{\circ}$.

\vspace{2cm}
\noindent 
DIAS-STP-97-24

\newpage

\noindent \section{{\bf Introduction}}
\vspace{.5cm}

$\sigma$-models in low dimensions have become an increasingly
important area of research. In two Euclidean dimensions they
appear to be the low dimensional analogues of four-dimensional
Yang-Mills theories. They often arise as approximate models,
in the context of both particle and solid state physics. They
have recently been used in the construction of models of high $T_c$ 
superconductivity \cite{one:1} and of the quantum Hall effect \cite{two:2}. 
Moreover, they are simple examples of harmonic maps studied
by differential geometers and, as such, are interesting in 
themselves. In addition, because of the nonlinearity of these
models, their classical solutions represent structures which
resemble solitons of (1+1) dimensional models and could 
become very useful in the description of many physical 
phenomena.

Recently, interesting scattering processes of these soliton-like 
objects were studied with the help of the collective coordinates 
approximation \cite{thr:3}\cite{fou:4}\cite{fiv:5} and with the help of 
numerical simulations. The numerical simulations were performed for the 
$CP^1$ model \cite{six:6}\cite{sev:7}, for its modification by the addition
of potential-like and Skyrme-like terms \cite{eig:8}\cite{nin:9},
and most recently for the $CP^2$ model with and without a generalized 
Hopf term \cite{ten:10}\cite{ele:11}. In Ref. 11, also the Ginibre et al. 
existence proof \cite{twe:12} and the Cauchy-Kowalewskyi theorem were used 
to study $CP^2$ soliton scattering. Although there is no mathematically 
rigorous justification for the collective coordinate approximation
in the $CP^n$ model yet, Manton's arguments, first put 
forward for the monopole motion \cite{thi:13}, should also apply to the 
$CP^2$
soliton motion. In this paper we therefore use this approximation to 
study some aspects of the $CP^2$ soliton scattering. We again find 
the $90^{\circ}$ scattering
without, and a deviation from the $90^{\circ}$ scattering, with the generalized 
Hopf term. 

The paper is organized as follows. In section 2, we describe the
$CP^2$ model without and with generalized Hopf term. We also 
formulate a Cauchy problem suitable for the description of solitons 
which merge at some stage of the scattering process. 
In section 3, the description of the scattering process in terms of
collective coordinates is given. In section 4, we solve the equations
for the collective coordinates for solitons close enough together, i.e. 
for short enough times before and after the collision.
Solitons far apart are studied in section 5.

\vspace{1cm}

\noindent \section{{\bf The model}}
\vspace{.5cm}

The Lagrangian for the $CP^2$ model is,
\begin{equation}
{\cal L}_{0} = (D^{\mu} Z^a)^{*} (D_{\mu} Z_a), \;\;\;\;\;\;\;\; \mu = 0,1,2;
\;\; a = 1,2,3;
\end{equation}
where $Z_a$ is a complex function of $(t,x^1,x^2)$ on $M^3$ (or a
function of the variables $(t,z={1 \over 2} (x^1 + i x^2),z^{*})$),
and $D_{\mu} Z_a = \partial_{\mu} Z_a - Z_{b}^{*}(\partial_{\mu} Z^{b})
Z_a$. Also, the $Z_a$ have to satisfy the condition $Z_{a}^{*} Z^{a}=1$.
$a,b,..$ are raised and lowered with the metric $\eta = diag(+1,+1,+1)$.
The space-time metric is $g = diag(+1,-1,-1)$.
The equations of motion corresponding to (2.1) are,
\begin{equation}
D_{\mu} D^{\mu} Z_a + (D_{\mu} Z_b)^{*} (D^{\mu} Z^b) Z_a = 0,
\end{equation}
together with $Z_{a}^{*} Z^a =1$. In this paper,
we will concentrate on solutions with winding number $k$, defined by
\begin{equation}
k = \frac{\imath}{2\pi} \int_{{\bf R}^2} \epsilon_{jk} 
(D^j Z^a)^* (D^k Z_a) \; d^{2}x,
\end{equation}
equal to 2.

The initial data are calculated from the complex functions,
\[
w_{1}(0,z,z^{*}) = \lambda z^2, \;\;\; w_{2}(0,z,z^{*}) = \beta z,
\]
\begin{equation}
\partial_{0} w_{1}(0,z,z^{*}) =  \lambda v, \;\;\;
\partial_{0} w_{2}(0,z,z^{*}) = 0,
\end{equation}
where $\lambda, \beta$ and $v$ are real positive constants,
and where $w_{i}(t,z,z^{*})$ and $Z_a$ are related through
\begin{equation}
Z = \frac{1}{\sqrt{1+|w_{1}|^2+|w_{2}|^2}} \left(
\begin{array}{c}
1 \\ w_{1} \\ w_{2}
\end{array} \right).
\end{equation}
For these initial data the energy is finite and the winding number
is $2$. The initial data describe
a head-on collision of solitons which 'coincide' at $t=0$. 
Negative time is the time before, and positive time is the time
after the collision.
The condition, $|w_{1}|^2+|w_{2}|^2=0$ at $z=0$, which holds 
for the initial data, reflects the fact that the solitons
merge at $t=0$. 

For $\partial_{0} w_{i}$ at $t=0$, 
we have taken zero modes which give a head-on collision. This means
that we don't need any extra potential energy to go 
through the 'ring' which forms at $t=0$. For small $v$, 
there is hardly any excess energy, and the slow motion approximation, or 
collective coordinate approximation, can be used. The idea of this 
approximation \cite{thi:13} is to describe the solution at each fixed time
in terms of a configuration with minimal potential energy. Then the action 
is minimized to obtain the collective coordinates, which parameterize 
these minimal-energy configurations, as functions of time.

For the $CP^2$ model the configurations with minimal potential energy,
which for constant parameters are time independent solutions, are of the
form (2.5) where $w_1$ and $w_2$ are rational functions of 
$z$ \cite{fou:14}. These configurations can also be given in the form,
\begin{equation}
Z_a = \frac{f_a}{|f|} \;\;\; \hbox{with} \;\;\; |f| = \sqrt{f_a^* f^a},
\end{equation}
where the $f_a$'s are polynomials in $z$. To relate the two descriptions
use is made of the $U(1)$ symmetry of the theory. In terms of $f_a$, the
equations of motion (2.2) read
\begin{equation}
\partial_{\mu} \partial^{\mu} f^a - \frac{f^a f^*_b \partial_{\mu}
\partial^{\mu} f^b}{|f|^2} = \frac{2(\partial_{\mu} f^a)
f^*_b \partial^{\mu} f^b}{|f|^2} - \frac{2 f^a f^*_b
(\partial_{\mu} f^b) f^*_c \partial^{\mu} f^c}{|f|^4}.
\end{equation}
The winding number can be expressed as
\begin{equation}
k = \frac{1}{2\pi} \int_{0}^{2\pi} \lim_{r\rightarrow\infty}
\frac{f^*_a z \partial_z f^a}{|f|^2} d\theta,
\end{equation}
where $z= r e^{\imath \theta}$. This formula shows that for $k=2$,
we can restrict our attention to quadratic polynomials $f_a$.

In sections 4 and 5, we will also study the effect the generalized 
Hopf term,
\begin{equation}
{\cal L}_{Hopf} = \epsilon^{\mu\nu\rho}
(D_{\mu}Z^{a})^{*} (D_{\nu}Z_{a})
(Z^{b*}\partial_{\rho}Z_{b}),
\end{equation}
has on the scattering process. For the Lagrangian 
${\cal L}_{0} + \kappa {\cal L}_{Hopf}$, the equations of
motion read,
\begin{equation}
(\delta^a_b -Z_{a}Z^{b*})D_{\mu}D^{\mu}Z_{b} - 2\kappa\epsilon^{\mu\nu\rho}
D_{\rho}Z_{a}((D_{\mu}Z^{b})^{*}D_{\nu}Z_{b}) = 0,
\end{equation}
where, as before, $Z^{*}_{a}Z^{a}=1$.

\vspace{1cm}

\noindent\section{{\bf The collective coordinate ansatz}}
\setcounter{equation}{0}
\vspace{.5cm}

For the collective coordinate approximation, we take the static solutions,
$Z_a(z,c_r)$, parameterized by parameters $c_r$, make the parameters time 
dependent and minimize within this ansatz. If we work with configurations
of the form (2.6), this means that
\begin{equation}
f_a (t,z) = \sum_{l=0}\sp{n}c^{n}_{a} (t) z^n,
\end{equation}
where $n=2$, for a 2-soliton configuration. 

Within the ansatz, the
potential energy is a constant and the $CP^2$ kinetic energy is
\[
T_0 = (D^0 Z^a )^{*} (D_0 Z_a )
\]
\begin{equation}
= \frac{1}{|f|^2 } 
((\partial_0 f^{*}_a )(\partial^0 f^a ) - \frac{1}{|f|^2 }
f^{*}_a (\partial_0 f^a ) f^b (\partial^0 f^{*}_b )).
\end{equation}
This yields,
\begin{equation}
\int_{{\bf R}^2 } T_0 d^{2}x = A^{mn}_{ab} \dot{c}^{m*}_{a} \dot{c}^{n}_{b} = L
\end{equation}
with,
\begin{equation}
A^{mn}_{ab} = \int_{{\bf R}^2 } \frac{1}{|f|^2} (\delta_{ab} z^{*m} z^n
- \frac{1}{|f|^2} c^{r}_{a} c^{s*}_{b} z^{n+r} z^{*m+s}) d^{2}x.
\end{equation}
 
Before we minimize the action, $\int_{-\infty}^{\infty} L dt$, in the 
next two sections, we will simplify the ansatz (3.1).
Our next arguments work for all initial data, including ours in (2.4), 
which satisfy the following conditions:
$f^a = f^a (z), a=1,2,3$, with $\partial_z f^a \neq 0$ for $a=2,3$; 
$\partial_0 f^k = 0, k=1,3$; $\partial_0 f^2
= \partial_0 f^2 (z) \neq 0$. For $\kappa =0$, we find
$f^1 \partial^{2}_{0} f^3 - f^3 \partial^{2}_{0} f^1 = 0$ 
at $t=0$. To derive this result, eq. (2.7), the conditions on the initial 
data and the formulas $-\partial_i \partial^i = \partial_z \partial_{z^*}$
and $-2 (\partial_i f^a) \partial^i f^b = (\partial_z f^a)
\partial_{z^*} f^b + (\partial_{z^*} f^a) \partial_z f^b$  
for $a,b=1,2,3$ are used.
Now $f^1$ can be made real by a gauge transformation, and multiplying
all $f^a$'s by $1/f^1$ we can achieve $f^1=1$ for all $t$. So at $t=0$ we 
have $\partial_{0}^2 f^3 =0 $. We show next that
$\partial_{0}^3 f^3 =0$ for $t=0$. For this we need again that the initial data
depend on $z$ only. 

To go further we need next that $\partial_0^2 f^2$
at $t=0$ depends on $z$ only. This is not the case for the exact solution. 
In fact we obtain
\begin{equation}
\partial_0^2 f^2 = \frac{2}{|f|^2} f^*_2 (\partial_0 f^2)^2
\end{equation}
at $t=0$. However, within the confines of the collective coordinate
approximation, we are neglecting contributions due to a 
$z^*$ dependence of $f_a$, and in that case we can 
actually show that all time derivatives of $f^3$ vanish 
at $t=0$. The reasonable assumption of analyticity now leads to the
time independence of $f^3$ for $t>0$.

So far we have $f^1 =1$, $f^3 = \beta z$ and  
\begin{equation}
f^2 = \frac{a_1 (t) z^2 + a_2 (t) z + a_3 (t)}{a_4 (t) z + a_5 (t)}
\end{equation}
If $a_4$ depends on time the kinetic energy, $\int_{{\bf R}^2} T_0 d^2 x$, 
diverges. (The interpretation of this divergence is that a
change in $a_4$ costs infinite energy.) So $a_4$ must be constant,
and therefore zero because of the initial data. The kinetic energy also 
diverges for time dependent $a_1$. Hence we have $a_1 = \lambda$
and $f^2 = \lambda z^2 + \delta (t) z + \gamma (t)$. Next we want
to show that $\delta$ can be taken to be zero.

The Euler-Lagrange equations for $L$ defined in eq. (3.3), the 
geodesic equations, are,
\begin{equation}
A^{mn}_{ab} \ddot{c}^{n}_{b} 
+ \frac{\partial A^{mn}_{ab}}{\partial c^{r}_{c}}
\dot{c}^{r}_{c} \dot{c}^{n}_{b}
+ \frac{\partial A^{mn}_{ab}}{\partial c^{r*}_{c}}
\dot{c}^{r*}_{c} \dot{c}^{n}_{b}
- \frac{\partial A^{rn}_{cb}}{\partial c^{m*}_{a}} 
\dot{c}^{r*}_{c} \dot{c}^{n}_{b} = 0,
\end{equation}
with $c^n_a (t)$ defined in (3.1).
If $c^{0}_{2} (t) = \gamma (t)$ and the other $c^{n}_{a}$ are constant, 
then the equations reduce to,
\begin{equation}
A^{m0}_{a2} \ddot\gamma + \frac{\partial A^{m0}_{a2}}{\partial \gamma}
\dot\gamma^2 = 0.
\end{equation}
Here we have used that,
\begin{equation}
\frac{\partial A^{m0}_{a2}}{\partial c^{0*}_{2}} =
\frac{\partial A^{00}_{22}}{\partial c^{m*}_{a}}.
\end{equation}
That (3.9) holds can be seen by taking the corresponding derivatives 
of the integrands which define the $A$'s in eq. (3.4).

If on the other hand we first set $c^{0}_{2} = \gamma (t)$ and the 
other $c^{n}_{a}$ constant in eq. (3.3), 
and then minimize, we obtain,
\begin{equation}
A^{00}_{22} \ddot\gamma + 
\frac{\partial A^{00}_{22}}{\partial \gamma} \dot\gamma^2 = 0.
\end{equation}
The equations (8) are now satisfied if,
\begin{equation}
A^{m0}_{a2} \frac{\partial A^{00}_{22}}{\partial \gamma}
= A^{00}_{22} \frac{\partial A^{m0}_{a2}}{\partial \gamma}.
\end{equation}
In our case, so far the ansatz has been reduced to $f^1 =1$, 
$f^2 = \lambda z^2 + \delta (t) z +\gamma (t)$, $f^3 = \beta z$.
For this ansatz, we have to solve the geodesic equations (3.7). A special 
solution is provided by $\delta =0$, which satisfies the initial 
condition, and a function
$\gamma$ which satisfies (3.10). Now (3.11) holds trivially,
since both the integrand of $A^{10}_{22}$ and the integrand of
$\partial A^{10}_{22} / \partial\gamma$ have the symmetry
$I(\theta + \pi) = -I(\theta)$, as functions of the angle
$\theta$. Hence the corresponding integrals vanish.
Therefore, finally our collective coordinate ansatz is 
\begin{equation}
f^1 = 1, \;\; f^2 = \lambda z^2 + \gamma (t), \;\; f^3 = \beta z.
\end{equation}

If the Hopf term is included, the Lagrange function, defined in (3.3), will
aquire the extra term $\kappa \int_{{\bf R}^2} {\cal L}_{Hopf} d^2 x$,
where
\[
{\cal L}_{Hopf} = \epsilon^{\mu \nu \rho} (\partial_{\mu} Z^{*}_{a})
(\partial_{\nu} Z^{a}) Z^{*}_{b} \partial_{\rho} Z^{b}
\]
\begin{equation}
= \frac{\epsilon^{ij}}{2|f|^4} ((\partial_i f^{*}_{a})
(\partial_j f^a ) (f^{*}_{b} \partial_0 f^b - f^b \partial_0 f^{*}_b )
\end{equation}
\[
+((\partial_0 f^{*}_a ) (\partial_i f^a ) 
-(\partial_i f^{*}_a )
(\partial_0 f^a )) (f^{*}_b \partial_j f^b - f^b \partial_j f^{*}_b )).
\]
We will study the effect of the Hopf term for $\kappa$ small enough so that 
the ansatz (3.12) still gives a reasonable approximation. If $\kappa$ is not
small, we do not expect a time independent $f_3$ to be a good 
approximation because
\begin{equation}
\partial^2_0 f_3 = \frac{-\imath \kappa \beta v |\lambda|^2 z^*
(2 + |\beta|^2 |z|^2 )}{(1+|\beta|^2 |z|^2 + |\lambda|^2 |z|^4)^2}
\end{equation}
holds for our initial data at $t=0$.

\vspace{1cm}

\noindent \section{{\bf Approximate solution for solitons close together}}
\setcounter{equation}{0}
\vspace{.5cm}

The ansatz (3.12) leads to
\begin{equation}
T_0 = \frac{1}{|f|^2} (1- \frac{|\lambda z^2 + \gamma |^2}{|f|^2})
\dot\gamma  \dot\gamma^{*} .
\end{equation}
Note that $T_0$ is a rational function of $r$, where 
 $z=r e^{\imath \theta}$, and that the
denominator is a product of quadratic functions in $r^2$. So the 
$r$ integration can be done, leading to functions
of $\theta$ which have to be integrated numerically. We also obtain,
\begin{equation}
{\cal L}_{Hopf} = \frac{i|\beta|^2}{4|f|^4}
((\gamma^{*} - \lambda^{*} z^{*2}) \dot \gamma
- (\gamma - \lambda z^2) \dot \gamma ^{*})
\end{equation}
Again this is a rational function of $r$ which can be integrated
over $r$.

In this section, we concentrate on solitons close together, i.e., we
choose $\vert\gamma\vert$ to be small.
For small $\vert\gamma\vert$, we expand,
\begin{equation}
\frac{1}{|f|^2} = \frac{1}{1+|\beta|^2  r^2 + |\lambda|^2 r^4}
- \frac{\lambda \gamma^{*} z^2 + \lambda^{*} \gamma z^{*2}} 
{(1+|\beta|^2 r^2 + |\lambda|^2 r^4)^2} + O(\vert\gamma\vert^2).
\end{equation}
Now we obtain,
\begin{equation}
\int_{{\bf R}^2} T_0 d^2 x = A\dot\gamma \dot\gamma^{*} 
+ O(\vert\gamma\vert^3),
\end{equation}
where
\begin{equation}
A = 2\pi \int^{\infty}_{0} dr \frac{r (1+|\beta|^2 r^2 )}
{(1+|\beta|^2 r^2 + |\lambda|^2 r^4 )^2}.
\end{equation}
The integral can be evaluated and given in terms of
\begin{equation}
g(\lambda ,\beta) = \left\{
\begin{array}{c}
\frac{2}{\sqrt{|\beta|^4 - 4|\lambda|^2}} \log (\frac{|\beta|^2}{2|\lambda|}
 + \frac{\sqrt{|\beta|^4 - 4|\lambda|^2}}{2|\lambda|} ) 
 \;\;\;\; for \;\;\;\; 2|\lambda| < |\beta|^2  \\
\frac{1}{|\lambda|} \;\;\;\;\;\;\; for \;\;\;\;\;\;\; 2|\lambda| = |\beta|^2  \\
\frac{2}{\sqrt{4|\lambda|^2 - |\beta|^4}} (\frac{\pi}{2} - 
\arctan \frac{|\beta|^2}{\sqrt{4|\lambda|^2 - |\beta|^4}} )
 \;\;\;\; for \;\;\;\; 2|\lambda| > |\beta|^2 
\end{array} \right.
\end{equation}
This yields,
\begin{equation}
A = \frac{\pi}{4|\lambda|^2 - |\beta|^4} 
[|\beta|^2 - (|\beta|^4 - 2 |\lambda|^2) g(\lambda ,\beta )]
\end{equation}
for $|\beta|^2 \neq 2|\lambda|$. For $|\beta|^2 = 2|\lambda|$, 
the formula reduces to $A=2\pi/(3|\lambda|)$.

We also have,
\begin{equation}
\int_{{\bf R}^2} {\cal L}_{Hopf} d^2x = \imath B (\gamma^{*}\dot\gamma
- \gamma \dot\gamma^{*}) + O(\vert\gamma\vert^3),
\end{equation}
where
\begin{equation}
B = \frac{\pi |\beta|^2}{2} \int_{0}^{\infty} dr
\left(\frac{r}{(1+|\beta|^2 r^2 + |\lambda|^2 r^4 )^2}
+ \frac{|\lambda|^2 r^5}{(1+ |\beta|^2 r^2 + |\lambda|^2 r^4 )^3}\right). 
\end{equation}
Note that $\beta = 0$ ($CP^1$ embedding) implies $B=0$.
Again the integral can be evaluated, and we obtain
\begin{equation}
B = \frac{\pi |\beta|^2}{4(4|\lambda|^2 - |\beta|^4)^2}
[|\beta|^2 (|\beta|^4 - 7 |\lambda|^2) + 
|\lambda|^2 (10 |\lambda|^2 - |\beta|^4) g(\lambda ,\beta )] 
\end{equation}
for $|\beta|^2 \neq 2|\lambda|$.  If $|\beta|^2 =2|\lambda|$, the result
reduces to $B=11\pi/60$.

For small $\vert\gamma\vert$, we end up with the Lagrange function,
\begin{equation}
L = A \dot\gamma \dot\gamma^{*} + \imath \kappa B (\gamma^{*} \dot\gamma 
- \gamma \dot\gamma^{*} ).
\end{equation}
Expressing $\gamma (t) = R(t) e^{\imath\sigma (t)}$ in terms of  
real functions, we obtain, 
\begin{equation}
L = A( \dot R ^2 + R^2 \dot\sigma ^2 ) - 2\kappa B R^2 \dot\sigma,
\end{equation}
and the equations of motion,
\begin{equation}
A \ddot R - A R \dot\sigma ^2 + 2 \kappa B R \dot\sigma = 0,
\end{equation}
\begin{equation}
A R^2 \ddot\sigma + 2AR \dot R \dot\sigma - 2 \kappa BR\dot R =0.
\end{equation}

The second equation implies,
\begin{equation}
\dot\sigma = \frac{c_1 + \kappa BR^2}{AR^2}.
\end{equation}
If $c_1 \neq 0$, $\dot\sigma \rightarrow \infty $ as $R \rightarrow 0$.
So we set $c_1 =0$, and obtain $\dot\sigma = \kappa B/A$ and
w.l.g. $\sigma = \kappa B t /A$. Now the equation for $R$ reads,
\begin{equation}
A^2 \ddot R + \kappa ^2 B^2 R = 0,
\end{equation}
and, using $R(0)=0$, we obtain,
\begin{equation}
R = c_2 \sin\frac{\kappa B t}{A} = c_2 \frac{\kappa B}{A} t,
\end{equation}
for small $R$ (and $t$).

We now consider $|f|$. At $t=0$, $|f|$ is radially symmetric
and describes a 'crater-like' surface. When $t$ becomes nonzero, the surface 
'buckles'. Consider, with $\lambda$ real for simplicity,
\begin{equation}
|f|^2 = 1 + |\beta|^2 r^2 + |\lambda|^2 r^4 + 2 \lambda r^2
R(t) \cos (2\theta - \sigma (t)).
\end{equation}
The $\theta$ derivatives gives us the 'ridges' and 'valleys',
which are at  $\theta = n\pi /2 + \sigma /2$. 
If for definiteness $\lambda > 0$ and $c_2 > 0$, the 'valleys' are at
$\frac{\kappa Bt}{2A}$ and $\pi +\frac{\kappa Bt}{2A}$ for $t<0$, and at
$\frac{\pi}{2} + \frac{\kappa Bt}{2A}$ and $\frac{3\pi}{2} + 
\frac{\kappa Bt}{2A}$ for $t>0$ shortly before and shortly after the
collision. For $\kappa = 0$, we again find $90^{\circ}$ scattering,
whereas for $\kappa\neq 0$, we find a deviation from this type
of scattering.

\vspace{1cm}

\noindent\section{{\bf Solitons at a distance}}
\setcounter{equation}{0}
\vspace{.5cm}

To study solitons which are not close together, $T_0$, from 
Eq.(4.1), and ${\cal L}_{Hopf}$, from Eq.(4.2), must be 
integrated over ${\bf R}^2$. For $T_0$,
the result of the $r$ integration can be written as 
\begin{equation}
\int_{{\bf R}^2} T_0 d^2 x = \int^{\infty}_{0} d\theta \;
a(\lambda ,\beta ,\gamma ,\theta) \; \dot\gamma \dot\gamma^{*}.
\end{equation}
$a(\lambda ,\beta ,\gamma ,\theta)$ can be expressed in terms of
\begin{equation}
b(\lambda ,\beta ,\gamma ,\theta) = |\beta|^2 + \lambda \gamma^* e^{2\imath\theta}
+ \lambda^* \gamma e^{-2\imath\theta},
\end{equation}
\begin{equation}
\Delta (\lambda ,\beta , \gamma,\theta) = 4|\lambda|^2 (1+|\gamma|^2) - b^2,
\end{equation}
and
\begin{equation}
h(\lambda ,\beta ,\gamma ,\theta) = \left\{
\begin{array}{c}
\frac{1}{\sqrt{-\Delta}} \log \frac{b+\sqrt{-\Delta}}{b-\sqrt{-\Delta}} 
 \;\;\;\; for \;\;\;\; \Delta < 0  \\
\frac{2}{b} \;\;\;\;\;\;\; for \;\;\;\;\;\;\; \Delta = 0  \\
\frac{2}{\sqrt{\Delta}} (\frac{\pi}{2} - 
\arctan \frac{b}{\sqrt{\Delta}} )
 \;\;\;\; for \;\;\;\; \Delta > 0 
\end{array} \right.
\end{equation}
as
\begin{equation}
a(\lambda ,\beta ,\gamma ,\theta) = \frac{|\beta|^2}{4|\lambda|^2 
(1+ |\gamma|^2)} + \frac{2|\lambda|^2 - |\beta|^2 b}{4|\lambda|^2\Delta}
\left[ 2|\lambda|^2 h(\lambda ,\beta ,\gamma ,\theta )
-\frac{b}{1+|\gamma|^2}\right].
\end{equation}
The $\theta$ integration and the integration of the Euler-Lagrange
equations have to be done numerically. 

For large $|\gamma|$ though, we can obtain analytical results. 
We start by integrating the factor of $\dot\gamma\dot\gamma^*$
on the right-hand side of (4.1) over regions close to $\pm a$. Set 
$\gamma = -\lambda a^2$ and $z=\pm a+b$ with $0\le\vert b \vert\le \vert a\vert 
^{3/4}$. (Here and in our arguments below, there is a wide choice of
limits of integration. Here for $\vert b\vert$, any power of
$\vert a\vert$ between $\frac{1}{2}$ and $1$ would do). Neglecting $b$ terms 
relative to $a$ terms, we get the integrals,
\begin{equation}
I_{\pm} = 4\pi \int_0^{\vert a\vert ^{3/2}}  \frac{1+\vert\beta\vert^2
\vert a \vert ^2}{(1+\vert\beta\vert^2\vert a\vert ^2 
+ 4\vert\lambda\vert ^2 \vert a\vert ^2 \vert b\vert^2)^2}
d\vert b\vert ^2.
\end{equation}
Hence,
\begin{equation}
I_{\pm} = \frac{\pi}{\vert\lambda\vert ^2 \vert a\vert ^2} - 
\frac{\pi (1+\vert\beta\vert^2\vert a\vert^2)}
{\vert\lambda\vert^2\vert a\vert^2 (1+\vert\beta\vert^2\vert a\vert^2 
+ 4\vert\lambda\vert^2\vert a\vert^{7/2})},
\end{equation}
and for large $\vert a\vert$,
\begin{equation}
I = I_+ + I_- = \frac{2\pi}{\vert\lambda\vert^2\vert a\vert^2}.
\end{equation}

We now show that the rest of the integral is of lower order than (5.8)
and can be neglected. This is an integral over ${\bf R}^2$ with two 
circles centered at $\pm a$ taken out. We split the integral into three
integrals, $I_1$, $I_2$ and $I_3$. $I_1$ is the integral for $\vert z\vert$ 
between 0 and $\vert a\vert^{3/4}$. $I_2$ results from the integration for 
$\vert z\vert$ from $\vert a\vert^{3/4}$ to $\vert a\vert ^2$, and $I_3$ is 
the integral over the rest of ${\bf R}^2$ outside the big circle of radius
$\vert a\vert ^2$. Starting with the integral $I_1$, we use the 
inequalities $\vert z \pm a\vert \ge\vert a\vert /2$, and take the lower 
bounds of these inequalities in the integrand. This yields, 
\[
I_1 \le 4\pi\int_0^{\vert a\vert^{3/2}} \frac{1+\vert\beta\vert^2
\vert z\vert^2}{(1+\vert\beta\vert^2\vert z\vert^2 + 
\vert\lambda\vert^2\vert a\vert^4 /16)^2} d\vert z\vert^2 
\]
\begin{equation}
\le 4\pi\int_0^
{\vert a\vert^{3/2}} \frac{1+\vert\beta\vert^2\vert z\vert^2}
{(1+\vert\lambda\vert^2\vert a\vert^4 /16)^2} d\vert z\vert^2.
\end{equation}
So the bound on $I_1$ is of order $1/ \vert a\vert^5$, and $I_1$ can 
be neglected.

The second integration area, which is an annulus with two small disks 
deleted, is split into two symmetric halves with a small disk deleted in 
each. In one half, $\vert z - a\vert \ge \vert a\vert^{3/4}$ 
and $\vert z + a\vert \ge \vert z\vert$; in the other,
$\vert z+a\vert \ge \vert a\vert ^{3/4}$ and $\vert z-a\vert \ge
\vert z\vert$. By taking the lower bounds of these inequalities in the
integrand, we get an upper bound for $I_2$. We now increase the value 
of this upper bound by including the two small disks. So we have
\[
I_2 \le 4\pi\int_{\vert a\vert^{3/2}}^{\vert a\vert ^4} \frac{1+\vert\beta
\vert^2 \vert z\vert^2}{(1+\vert\beta\vert^2\vert z\vert ^2 
+ \vert\lambda\vert ^2 \vert a\vert ^{3/2}
\vert z\vert ^2)^2} d\vert z\vert^2 = \frac{4\pi}
{(\vert\beta\vert^2 + \vert\lambda\vert ^2\vert a\vert ^{3/2})^2}
\]
\begin{equation} 
\times\left[\vert\beta\vert^2\log (1+\vert\beta\vert^2\vert z\vert^2 
+ \vert\lambda\vert^2 \vert a\vert^{3/2}\vert z\vert^2) 
- \frac{\vert\lambda\vert^2 \vert a\vert^{3/2}}{1+\vert\beta\vert^2
\vert z\vert^2 + \vert\lambda\vert^2\vert a\vert^{3/2}\vert z\vert^2} 
\right]_{\vert a\vert^{3/2}}^{\vert a\vert^4}.
\end{equation}
For large $\vert a\vert$, this goes like $\frac{\log\vert a\vert}
{\vert a\vert^3}$, and $I_2$ can be neglected. Finally,
for the integral $I_3$, we use the inequalities $\vert z \pm a\vert \ge 
\vert z\vert /2$. This leads to the following upper bound,
\begin{equation}
I_3 \le 4\pi\int_{\vert a\vert^4}^{\infty} \frac{1+\vert\beta\vert^2
\vert z\vert^2}{(1+\vert\beta\vert^2\vert z\vert^2 
+ \vert\lambda\vert^2\vert z\vert^4/16)^2} 
d\vert z\vert^2 \le \frac{2^{10}\pi\vert\beta\vert^2}
{\vert\lambda\vert^4 \vert a\vert^8}.
\end{equation}

To include the generalized Hopf term, we again perform the $r$ integration
and obtain
\begin{equation}
\int_{{\bf R}^2} d^2 x {\cal L}_{Hopf} = \frac{\imath |\beta|^2}{4}
\int^{2\pi}_{0} d\theta \; c(\lambda, \beta,\gamma,\dot\gamma,\theta)\; ,
\end{equation}
where
\[
c(\lambda,\beta,\gamma,\dot\gamma,\theta) =
\left[\gamma^* \dot\gamma - \gamma \dot\gamma^*
+\frac{b}{2|\lambda|^2} (\lambda^* \dot\gamma e^{-2\imath\theta}
- \lambda \dot\gamma^* e^{2\imath \theta})\right]
\]
\begin{equation}
\times\left[\frac{2|\lambda|^2}{\Delta} h(\lambda,\beta,\gamma,\theta)
-\frac{b}{\Delta (1+|\gamma|^2)}\right].
\end{equation}
Again, the $\theta$ integration and the integration of the Euler-Lagrange
equations have to be done numerically. 

For large $\vert a\vert$, however, we obtain some analytical results.
We must integrate ${\cal L}_{Hopf}$, given in (4.2), over ${\bf R}^2$. 
For large $\vert a\vert$, the leading term will again come from the 
integration over regions close to $\pm a$. As above we set $z=\pm a + b$ 
and integrate for $0< \vert b\vert^2 < \vert a\vert ^{3/2}$. Neglecting 
$b$ terms relative to $a$ terms, these integrals can be written as,
\begin{equation}
J_{\pm} = \int_0^{\vert a\vert ^{3/2}} \frac{4\pi\imath\vert \beta\vert ^2
\vert\lambda\vert ^2 \vert a\vert ^2 (a^* \dot{a} - a\dot{a}^* )}
{(1+\vert\beta\vert ^2 \vert a\vert ^2 + 4\vert\lambda\vert ^2
\vert a\vert ^2 \vert b\vert ^2 ) ^2} d\vert b\vert ^2.
\end{equation}
To leading order in $\vert a\vert$, we obtain,
\begin{equation}
J=J_+ + J_- =\frac{2\pi\imath}{\vert a\vert ^2} (a^*\dot{a} - a\dot{a}^*).
\end{equation}

We now show that the integral over ${\bf R}^2$, with the two disks
centered at $\pm a$ taken out, can be neglected compared to $J$. Again 
using the circles of radius $\vert z\vert = \vert a\vert ^{3/4}$ and 
$\vert z\vert = \vert a\vert ^2$, respectively, the integral is divided 
into three integrals, $J_1$, $J_2$ and $J_3$. With the help of the 
inequalities $\vert z\pm a\vert \ge \vert a\vert /2$ which we used above, 
we now get the bound,
\begin{equation}
\vert J_1\vert \le \frac{2^{11}\pi\vert\beta\vert ^2 \vert\dot{a}\vert}
{\vert\lambda\vert ^2 \vert a\vert ^{7/2}},
\end{equation}
on the integral over the disk at the origin. $J_1$ is clearly of lower
order than $J$. Integration over the annulus for $\vert a\vert^{3/2}
< \vert z\vert^2 < \vert a\vert^4$, with the two disks deleted, leads 
to the bound,
\[
\vert J_2\vert \le \frac{4\pi\vert\beta\vert ^2\vert\lambda\vert ^2
\vert a\vert \vert\dot{a}\vert}{(\vert\beta\vert ^2 + \vert\lambda\vert ^2
\vert a\vert ^{3/2})^2} 
\]
\begin{equation}
\times\left[ \log (1+\vert\beta\vert ^2 \vert z\vert ^2
+\vert\lambda\vert ^2 \vert a\vert ^{3/2} \vert z \vert ^2)
+ \frac{(1 - \vert a\vert ^2 \vert\beta\vert ^2 - \vert\lambda\vert^2
\vert a\vert^{7/2})}{1+\vert\beta\vert ^2 \vert z\vert ^2 + 
\vert\lambda\vert ^2\vert a\vert ^{3/2} \vert z\vert ^2} 
\right]^{\vert a\vert ^4}_{\vert a\vert ^{3/2}}.
\end{equation}
This bound is of order $\vert\dot{a}\vert / \vert a\vert ^{3/2}$, and
$J_2$ can be neglected. $J_3$ can be neglected as well, since
\begin{equation}
\vert J_3 \vert \le \frac{2^{10} \pi\vert\beta\vert ^2 \vert\dot{a}\vert}
{\vert\lambda\vert ^2 \vert a\vert ^7}.
\end{equation}

If we use $\vert\dot\gamma\vert^2 = 4\vert\lambda\vert^2\vert a\vert ^2
\vert\dot{a}\vert^2$ and eqs. (5.8) and (5.15), we find the following
Lagrange function for large $\vert a\vert$,
\begin{equation}
L = 8\pi \dot{a}\dot{a}^* + 2\pi\imath\kappa \left(\frac{\dot{a}}{a} - 
\frac{\dot{a}^*}{a^*}\right).
\end{equation}
The Hopf term is the time derivative of $2\pi\imath\kappa\ln\frac{a}{a^*}$
and does not contribute to the equations of motion. To leading
order in the distance between the solitons, we have $a(t) = a_0 + vt$.
This is the free motion seen in previous numerical studies.

\vspace{1cm}

\noindent \section{{\bf Conclusions}}
\vspace{.5cm}

In this paper, which a sequel to \cite{ele:11}, we continued our investigation
of the scattering of solitons in the $CP\sp2$ model in (2+1) dimensions.
In \cite{ele:11} we showed that the addition of an extra term, the ``generalised
Hopf term" changes dynamics of the solitons; the $90\sp{\circ}$ scattering
seen in head-on collisions of $CP\sp2$ solitons in the absence of this
term is replaced by a difflection, which for small values
of the coefficient of this term is proportional to this coefficient.
These expectations, based on the expansion of the field at the time
of overlap of the two solitons, were shown to be be born out by the results
of numerical simulations. The simulations have also shown that at larger
distances the generalised Hopf term seems to play little role;
all deflection of the solitons seems to be confined to the region when
they are very close together.

In this paper we have looked at this problem from the point of view of 
collective coordinates. We have argued that in this problem we can approximate
the field configuration by a simple polynomial expressions (with many terms
vanishing) and then looked at the effects of the nonvanishing coefficient
of the generalised Hopf term on the dynamics of these collective
coordinates. We have shown that this approximation is self consistent and that
it captures the main features of the data ({\it ie} the deflection from the
$90\sp{\circ}$ scattering beeing proportional to the coefficient
of the Hopf term). Suprisingly, the approximation also captures the effect
of the Hopf term becoming irrelevant (for classical dynamics) when the
solitons are well separated; in this case its contribution to the 
effective lagrangian becomes a total derivative. Moreover, the form of this
total derivative is very similar to what is seen in the $CP\sp1$ model
(though there this form is valid for all distances between solitons,
including the cases when they overlap). Thus we see that well separated
solitons in $CP\sp2$ model are not that different from solitons of the
$CP\sp1$ model; only when they are close together the nature of target manifold
begins to play an important role, and the generalised Hopf term 
brings out this difference.

In addition our work has given an extra support for the use of collective
coordinates; they provide a good approximation to the full dynamics
also when the generalised Hopf term is present.

\vspace{1cm}

\noindent {\large\bf Acknowledgements}
This work was partially supported by the grant BCA 96/024 of Forbairt
and the British Council.


\newpage

\begin{thebibliography}{99}

\bibitem {one:1}
Dzyaloshinskii I, Polyakov A and Wiegmann P 1988 Phys. Lett. A 
{\bf 127} 112

\bibitem {two:2}
Green A G, Kogan I I and Tsvelik A M 1996 Phys. Rev. B
{\bf 53} 6981

\bibitem {thr:3}
Ward R S 1985 Phys. Lett. B {\bf 158} 424

\bibitem {fou:4}
Leese R 1990 Nucl. Phys. B {\bf 344} 33

\bibitem {fiv:5}
Sutcliffe P M 1991 Nonlinearity {\bf 4} 1109 

\bibitem {six:6}
Leese R A, Peyrard M and Zakrzewski W J 1990 Nonlinearity 
{\bf 3} 387

\bibitem {sev:7}
Zakrzewski W J 1991 Nonlinearity {\bf 4} 429

\bibitem {eig:8}
Leese R A, Peyrard M and Zakrzewski W J 1990 Nonlinearity
{\bf 3} 773

\bibitem {nin:9}
Peyrard M, Piette B and Zakrzewski W J 1991 Nonlinearity
{\bf 5} 563

\bibitem {ten:10}
Piette B, Rashid M S and Zakrzewski W J 1993 Nonlinearity
{\bf 6} 1077

\bibitem {ele:11}
Burzlaff J and Zakrzewski W J 1996 Nonlinearity {\bf 9} 1317

\bibitem {twe:12}
Ginibre J and Velo G 1982 Ann. Phys. {\bf 142} 393

\bibitem {thi:13}
Manton N S 1982 Phys. Lett. B {\bf 110} 54

\bibitem {fou:14}
D'Adda A, L{\" u}scher M and Di Vecchia P 1978 Nucl. Phys. B
{\bf 146} 63

\end{thebibliography}
\end{document}